\documentclass[onecolumn,showpacs,preprintnumbers,elsart]{revtex4}

\usepackage{amsmath}
\usepackage{mathrsfs}
\usepackage{graphicx}
\usepackage{dcolumn}
\usepackage{bm}
\usepackage[center]{subfigure}
\usepackage{color}

\begin{document}

\title{Two-component solitons under a spatially modulated linear coupling:
Inverted photonic crystals and fused couplers}
\author{Yongyao Li$^{1,2}$}
\email{yongyaoli@gmail.com}
\author{Wei Pang$^{3}$}
\author{Shenhe Fu$^{4}$}
\author{Boris A. Malomed$^{2}$}
\email{malomed@post.tau.ac.il}
\affiliation{$^{1}$Department of Applied Physics, South China Agricultural University,
Guangzhou 510642, China \\
$^{2}$Department of Physical Electronics, School of Electrical Engineering,
Faculty of Engineering, Tel Aviv University, Tel Aviv 69978, Israel\\
$^{3}$ Department of Experiment Teaching, Guangdong University of
Technology, Guangzhou 510006, China. \\
$^{4}$State Key Laboratory of Optoelectronic Materials and Technologies,\\
Sun Yat-sen University, Guangzhou 510275, China}

\begin{abstract}
We study two-component solitons and their symmetry-breaking bifurcations
(SBBs) in linearly coupled photonic systems with a spatially inhomogeneous
strength of the coupling. One system models an \textit{inverted} virtual
photonic crystal, built by periodically doping the host medium with atoms
implementing the electromagnetically induced transparency (EIT). In this
system, two soliton-forming probe beams with different carrier frequencies
are mutually coupled by the EIT-induced effective linear interconversion.
The system is described by coupled nonlinear Schr\"{o}dinger (NLS) equations
for the probes, with the linear-coupling constant periodically modulated in
space according to the density distribution of the active atoms. The type of
the SBB changes from sub- to supercritical with the increase of the total
power of the probe beams, which does not occur in systems with constant
linear-coupling constants. Qualitatively similar results for the SBB of
two-component solitons are obtained, in an exact analytical form, in the
model of a fused dual-core waveguide, with the linear coupling concentrated
at a point.
\end{abstract}

\maketitle

%\pacs{42.65.Tg; 42.70.Qs;05.45.Yv}

%\email{yongyaoli@gmail.com}
%\email{stszjy@mail.sysu.edu.cn}

%\preprint{APS/123-QED}

% Force line breaks with \\



\section{Introduction}

The effect of the spontaneous symmetry breaking of solitons in two-component
linearly-coupled system was studied in detail in nonlinear optics \cite%
{Snyder} and Bose-Einstein condensates (BECs) \cite{Gubeskys,Trippenbach}.
The symmetry-breaking bifurcation (SBB) occurs as a transition of a
symmetric localized (solitonic) ground states into an asymmetric one when
the linear-coupling constant drops below a critical value. In optics, the
linear coupling originates from the overlapping of evanescent fields between
adjacent waveguides, such as in dual-core fibers \cite{Snyder}, \cite%
{Kivshar,Hung} and arrays of such fibers \cite{Amherst,Belgrade}, or from
the linear mixing of orthogonal polarizations induced by the twist or
elliptic deformation in bimodal fibers \cite{Malomed,Dror}. In binary BECs,
a similar effect originates from the interconversion between hyperfine
atomic states induced by a resonant electromagnetic wave, as demonstrated
theoretically in a variety of settings \cite{Ballagh}. There are two kinds
of the SBBs, sub- and supercritical. In the case of the subcritical symmetry
breaking (which is tantamount to the phase transition of the first kind),
the system features branches of asymmetric states which emerge as unstable
ones, going at first backward from the bifurcation point and undergoing the
stabilization after turning forward. In the case of the supercritical SBB
(it is tantamount to the phase transition of the second kind), asymmetric
branches emerge as stable ones and immediately go in the forward direction.
An interesting problem is a possibility to control the type of the symmetry
breaking, and thus to switch between the respective phase transitions of the
two kinds. For example, the addition of a periodic potential (optical
lattice) acting in the unconfined direction changes the character of the
SBB\ from sub- to supercritical \cite{Trippenbach}), and a similar effect is
induced by rendering interactions nonlocal \cite{Brazhnyi}.

{One of versatile techniques for the control of transmission properties of
optical media is the electromagnetically induced transparency (EIT) \cite%
{LDeng}. The EIT gives rise to a variety of nonlinear features, which can be
used for the making of single- and multi-component solitons, and thus,
subsequently, for the implementation of the symmetry breaking in solitons.
These features include the self-enhanced \cite{XiaoM} and giant \cite%
{Schmidt} Kerr effects, as well as the enhanced frequency conversion \cite%
{Harris1990,Fleischhauer,HemmerHarrisHakuta,Korsunsky}.}

The first aim of the present work is to study the SBB for solitons in
two-component photonic systems featuring spatial modulations of the strength
of the linear coupling induced by the EIT-mediated frequency conversion. One
such system represents a virtual photonic crystal (PhC) formed by a periodic
modulation of the concentration of active atoms doping a passive host
medium. This technique has been recently implemented in the fabrication of
\textit{imaginary-part PhCs} by implanting atoms of RhB (Rhodamine B, a dye
manifesting saturable absorption) into the SU-8 polymer, which is a commonly
used transparent negative photoresist \cite{LJT}. {The present model refers
to another pair of the active atoms and passive background, namely, Pr$^{3+}$
ions and the YSO crystal, respectively. }

\begin{figure}[tbp]
%并排插入两个子图形
\centering
\subfigure[]{ \label{fig_1 a}
\includegraphics[scale=0.25]{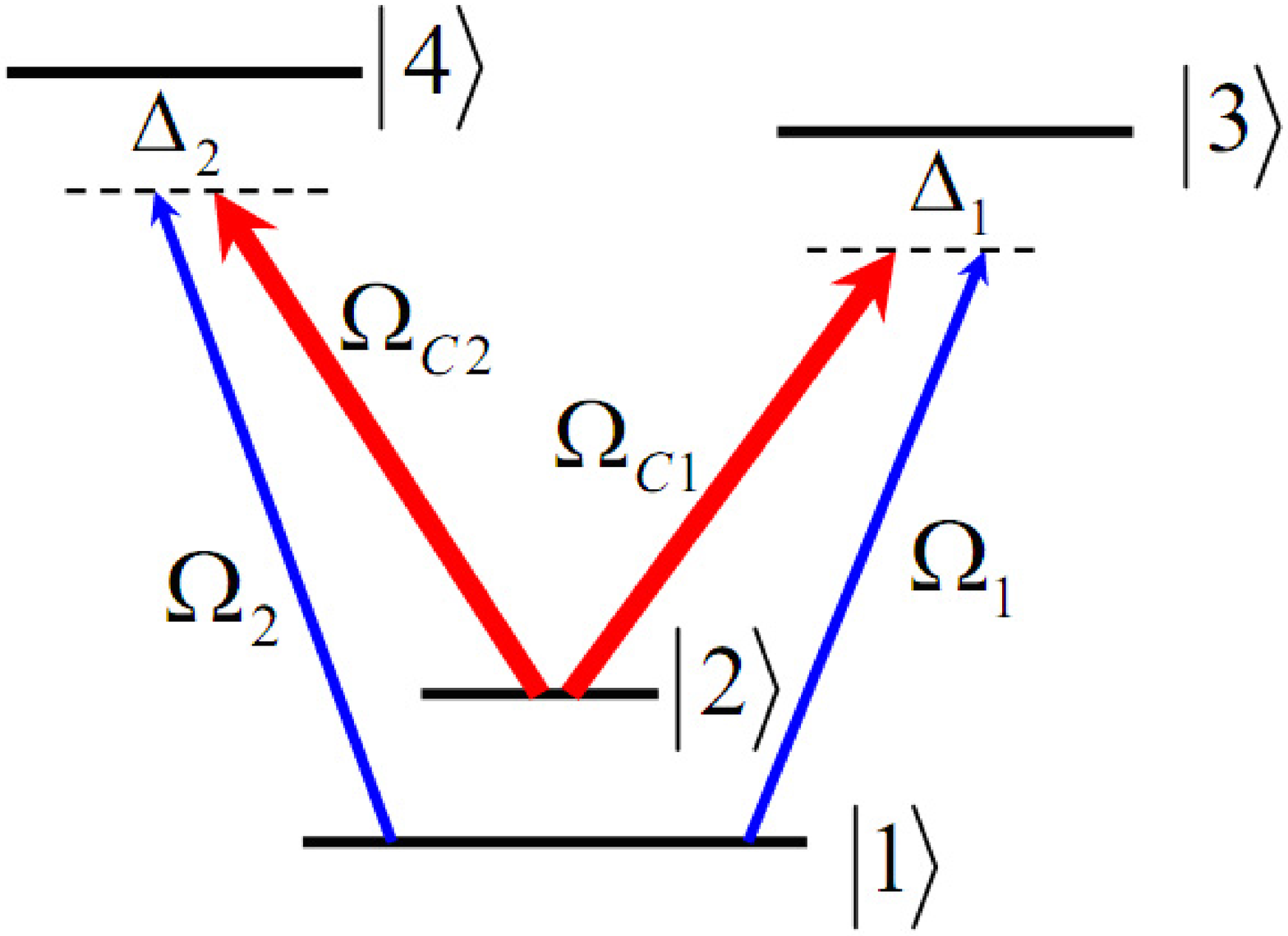}} \hspace{0.02in}
\subfigure[]{
\label{fig_1_b} \includegraphics[scale=0.16]{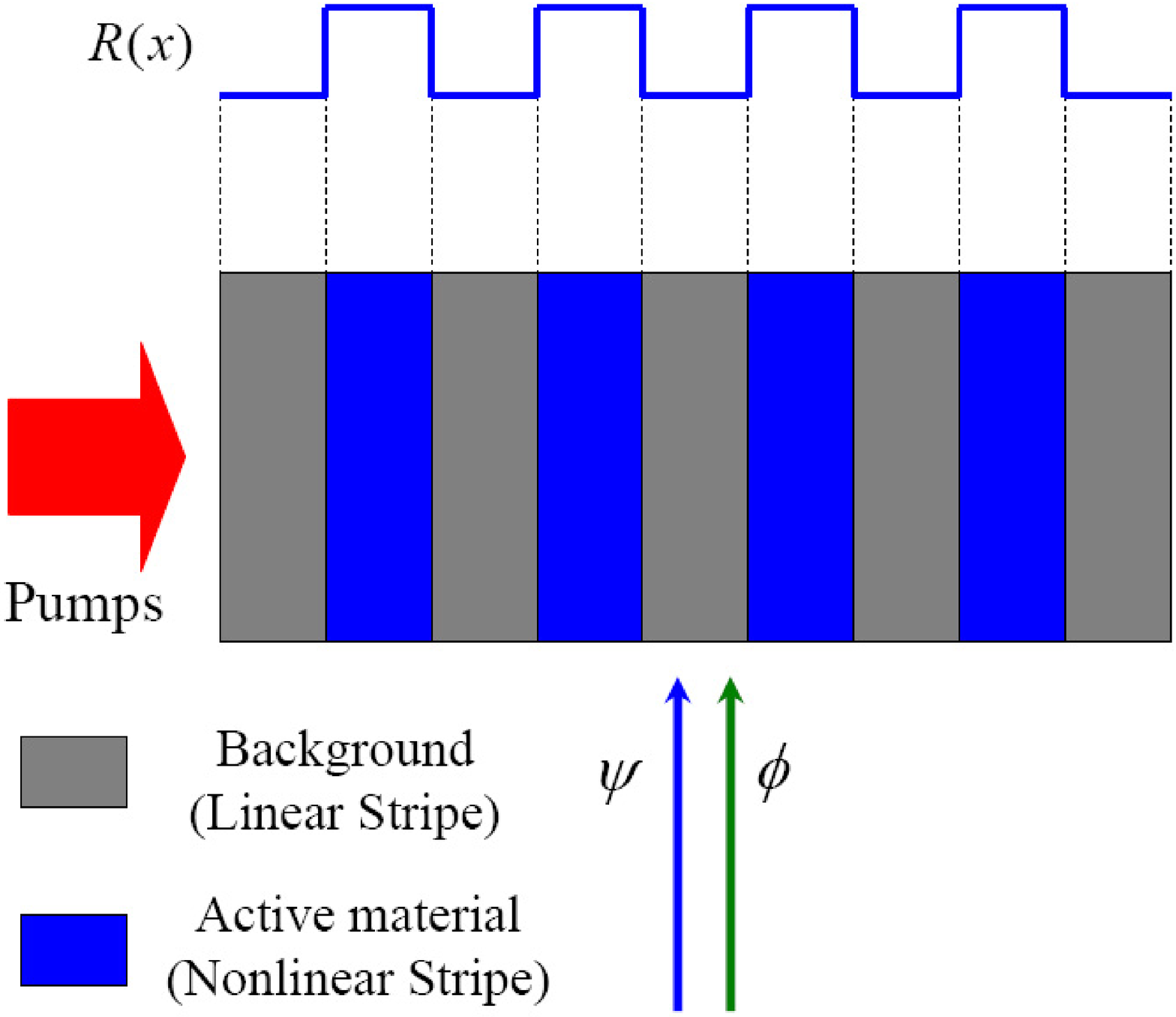}}
\caption{(Color online) (a) The energy-level structure (the double-$\Lambda $
scheme) of the active dopants (Pr$^{3+}$). (b) {The periodic distribution of
the dopants (filling blue regions) against the background (gray regions).
The dopant density is represented by function $R(x)$}}
\label{fig1}
\end{figure}

The energy scheme of Pr$^{3+}$ is displayed in Fig. \ref{fig1}(a), which is
a four-level energy system. Fields $\Omega _{1}$ and $\Omega _{2}$, which
have different carrier frequencies, represent two soliton-forming beams.
They induce resonant transitions from the same ground state, i.e., $%
|1\rangle $, to different excited states, $|3\rangle $ and $|4\rangle $,
with detuning $\Delta _{1}$ and $\Delta _{2}$, respectively. The resonant
transitions from a common metastable state, $|2\rangle $, to the same pair
of the excited states ($|3\rangle $ and $|4\rangle $) are powered by pump
(coupling) fields $\Omega _{C1}$ and $\Omega _{C2}$, with detunings adjusted
so that $\Delta _{C1}=\Delta _{1}$ and $\Delta _{C2}=\Delta _{2}$, which
nullifies the two-photon detuning. The four probe and pump fields involved
into the scheme build a double-$\Lambda $ scheme \cite{Korsunsky}. {In this
setting, the transitions driven by coupling fields }$\Omega _{C1}$ and $%
\Omega _{C2}${\ give rise to the effective linear mixing between fields $%
\psi $ and $\phi $, in the form of the \emph{induced frequency conversion}
\cite{HemmerHarrisHakuta,Korsunsky}. The latter feature places the system
into the class of the linearly-coupled two-component ones. }
{\ }

The periodic modulation of the dopant concentration, {i.e., the structure
function $R(x)$}, which represents the density of the implanted ions in Fig. %
\ref{fig1}(b), turns the uniform medium into a \textit{virtual} PhC \cite%
{LYY1}-\cite{Jianxiong}, so called because its properties are controlled by
the pump fields, rather than by a permanent material structure (other
well-known examples of virtual structures are lattices optically induced in
photorefractive crystals \cite{Moti}). The density-modulation pattern
induces an effective periodic linear potential, along with a spatially
periodic modulation of all local optical parameters, including the strength
of the linear interconversion and coefficients of the self-phase modulation
(SPM). The virtual PhC of this type may be naturally called an \textit{%
inverted} one, in comparison with ordinary (material) PhCs built in
self-focusing media: While the ordinary PhC structures feature effective
linear and nonlinear potentials with coinciding local minima, in the present
setting minima of the local potential coincide with maxima of its nonlinear
counterpart, and vice versa. i.e., the nonlinear potential is inverted with
respect to the linear one \cite{LYY1}-\cite{LYY3}. Similar models with
competing effective linear and nonlinear potentials were introduced, in a
different context, in Ref. \cite{Kar12}. In fact, this setting may be
considered as a variety of the general concept of \textit{nonlinear lattices}
and mixed linear-nonlinear ones \cite{NL}.

The studies of the single-component model of the inverted virtual PhC with
the Kerr and saturable nonlinearities, reported in Refs. \cite{LYY1} and
\cite{LYY2}, respectively, have revealed that the stable location of the
solitons in such systems is controlled by the total power, with the low- and
high-power solitons tending to be pinned by minima of the linear and
nonlinear potentials, respectively. Transitions between the different
positions of the solitons were identified as supercritical SBBs.

The major objective of this work is to study two-component solitons,
supported by the EIT scheme, in the inverted crystals, the main point being
the SBBs of the solitons. In Sec. II, we derive the model by means of the
semi-classical consideration of the interaction of the electromagnetic
fields with atoms. Then, we focus on effects of the periodically modulated
linear coupling between the two components, which is the main novel element
of the model. In Sec. III, we consider the competition of the linear
coupling and SPM nonlinearity. We concentrate on the SBB specific to the
two-component system, and do not dwell on the above-mentioned breaking of
the spatial symmetry of the solitons, which can be adequately studied in the
single-component system \cite{LYY2,LYY3}. A systematic numerical analysis
demonstrates a transition of the subcritical SBB into a supercritical one
with the increase of the soliton's total power, and a concomitant shrinkage
of the soliton. In Sec. IV, analytical results are presented for an allied
model of a dual-core \textit{fused} \cite{fused} spatial coupler, with the
tightly concentrated linear coupling represented by the delta-function (in
fact, the same system, but with a periodic array of coupled sites, and a
periodic modulation of the refractive index and Kerr coefficient, may also
represent the model of the virtual PhC considered in Sec. II). In that
model, the SBB of two-component solitons can be studied in an exact
analytical form. The paper is concluded by Sec. V.

\section{The model of the inverted nonlinear photonic crystal}

The atomic Hamiltonian corresponding to the configuration shown in Fig. \ref%
{fig_1 a} is
\begin{eqnarray}
H &=&\hbar \lbrack \delta |2\rangle \langle 2|+\Delta _{1}|3\rangle \langle
3|+\Delta _{2}|4\rangle \langle 4|]  \notag \\
&&-\hbar \lbrack \Omega _{1}|3\rangle \langle 1|+\Omega _{2}|4\rangle
\langle 1|+\Omega _{C1}|3\rangle \langle 2|+\Omega _{C2}|4\rangle \langle 2|+%
\mathrm{H.c.}],  \label{actomHami}
\end{eqnarray}%
where fields $\Omega $ and detunings $\Delta _{1,2}$ are defined as per the
figure, $\mathrm{H.c.}$stands for the Hermitian-conjugate contribution, and
%\begin{eqnarray}
%H=-\hbar\left(
%\begin{array}{cccc}
% 0 & 0 & \Omega^{\ast}_{1} & \Omega^{\ast}_{2} \\
% 0 & -\delta &  \Omega^{\ast}_{C1} & \Omega^{\ast}_{C1}\\
% \Omega_{1} & \Omega_{C1} & -\Delta_{1} & 0 \\
% \Omega_{2} & \Omega_{C2} & 0 & -\Delta_{2}
%\end{array}
%\right)
%\end{eqnarray}
the two-photon detuning, $\delta $, is set to be zero. Under physically
realistic conditions, spontaneous decay of the states included into the
scheme may be neglected. Further, if the dopant atoms are initially kept in
the ground state, steady-state solutions for density-matrix elements, $\rho
_{31}$ and $\rho _{41}$, which are activated by the two soliton-forming
probes, can be written as follows:
\begin{eqnarray}
&&\rho _{31}=-{\frac{\Delta _{2}}{\Delta _{1}}}{\frac{|\Omega _{1}|^{2}}{%
\Omega _{C}^{2}}}\Omega _{1}-{\frac{|\Omega _{2}|^{2}}{\Omega _{C}^{2}}}%
\Omega _{1}+{\frac{|\Omega _{C2}|^{2}}{\Omega _{C}^{2}}}\Omega _{1}-{\frac{%
\Omega _{C1}\Omega _{C2}^{\ast }}{\Omega _{C}^{2}}}\Omega _{2},  \notag \\
&&\rho _{41}=-{\frac{\Delta _{1}}{\Delta _{2}}}{\frac{|\Omega _{2}|^{2}}{%
\Omega _{C}^{2}}}\Omega _{2}-{\frac{|\Omega _{1}|^{2}}{\Omega _{C}^{2}}}%
\Omega _{2}+{\frac{|\Omega _{C1}|^{2}}{\Omega _{C}^{2}}}\Omega _{2}-{\frac{%
\Omega _{C2}\Omega _{C1}^{\ast }}{\Omega _{C}^{2}}}\Omega _{2},  \label{rho}
\end{eqnarray}%
where $\Omega _{C}^{2}\equiv \Delta _{2}|\Omega _{C1}|^{2}+\Delta
_{1}|\Omega _{C2}|^{2}$. The first (SPM)\ terms on the right-hand sides of
Eqs. (\ref{rho}) originate from the self-enhanced Kerr effect \cite{XiaoM},
the second and the third (XPM)\ terms represent the giant Kerr effect \cite%
{Schmidt}, and the fourth terms represent the linear coupling between the
soliton-forming beams, which originate from the EIT-induced frequency
conversion \cite{HemmerHarrisHakuta}.

The polarization experienced by the two probes in the medium are \cite%
{Scully}
\begin{eqnarray}
&&\mathscr{P}_{1}(x)=2N(x)\wp _{31}\rho _{31},  \notag \\
&&\mathscr{P}_{2}(x)=2N(x)\wp _{41}\rho _{41}.  \label{polar}
\end{eqnarray}%
Here,$\wp _{31}$ and $\wp _{41}$ (which are assumed real) are the matrix
elements of the dipole transitions $|1\rangle \rightarrow |3\rangle $ and $%
|1\rangle \rightarrow |4\rangle $. A reasonable simplification of the model
is attained by assuming that $\wp _{31}\approx \wp _{41}\equiv \wp $. The
paraxial propagation equations for the slowly varying envelopes of the probe
fields are
\begin{equation}
i\partial _{z}\Omega _{j}=-{\frac{1}{2k_{j}}}\partial _{xx}\Omega _{j}-{%
\frac{k_{j}\wp }{2\epsilon _{0}\hbar }}\mathscr{P}_{j}(x),[j=1,2].
\label{Maxeq}
\end{equation}%
Substituting Eq. (\ref{rho}) and Eq. (\ref{polar}) into Eqs. (\ref{Maxeq}),
one arrives at coupled nonlinear Schr\"{o}dinger (NLS) equations,
\begin{eqnarray}
&&i\partial _{z}\psi =-{\frac{1}{2}}\partial _{xx}\psi +R(x)\left( V_{1}\psi
+\sigma _{1}|\psi |^{2}\psi +\kappa |\phi |^{2}\psi +C_{1}\phi \right) ,
\notag \\
&&i\partial _{z}\phi =-{\frac{1}{2}}\partial _{xx}\phi +R(x)\left( V_{2}\phi
+\sigma _{2}|\phi |^{2}\phi +\kappa |\psi |^{2}\phi +C_{2}\psi \right) .
\label{twocomp}
\end{eqnarray}%
where $\psi =\Omega _{1}/\gamma $, $\phi =\Omega _{2}/\gamma $, $R(x)={\wp
^{2}}N(x)/\epsilon _{0}\hbar \gamma ^{2}$, $V_{1}=-|\Omega _{C2}|^{2}/\Omega
_{C}^{2}$, $V_{2}=-|\Omega _{C1}|^{2}/\Omega _{C}^{2}$, $\sigma _{1}=\Delta
_{2}/\Delta _{1}\Omega _{C}^{2}$, $\sigma _{2}=\Delta _{1}/\Delta _{2}\Omega
_{C}^{2}$, and $\kappa =1/\Omega _{C}^{2}$. If we let $\Omega _{C1}$ and $%
\Omega _{C2}$ be real, then coefficients $C_{1}=C_{2}\equiv C=\Omega
_{C1}\Omega _{C2}/\Omega _{C}^{2}$ account for the EIT-induced linear mixing
of the probe fields. The effective linear-coupling coefficient $CR(x)$ in
Eq. (\ref{twocomp}), which is periodically modulated due to the distribution
of the dopant density, makes the system different from various previously
studied models of linearly-coupled systems \cite{Snyder}-\cite{Amherst},
\cite{Dror}.

\section{Numerical results for the model of the photonic crystal}

To focus on effects of the periodically modulated linear coupling competing
with the SPM terms, we drop the XPM interaction in Eq. (5), and simplify
Eqs. (\ref{twocomp}) to the following form:
\begin{eqnarray}
&&i\partial _{z}\psi =-{\frac{1}{2}}\partial _{xx}\psi +V(x)\left( 1-|\psi
|^{2}\right) \psi -C(x)\phi ,  \notag \\
&&i\partial _{z}\phi =-{\frac{1}{2}}\partial _{xx}\phi +V(x)\left( 1-|\phi
|^{2}\right) \phi -C(x)\psi .  \label{eqs}
\end{eqnarray}%
These equations generalize those derived in Refs. \cite{LYY1,LYY2} for the
inverted PhC with the $\pi $-shift between the periodic linear and nonlinear
potentials, under the assumption that the depletion of the coupling fields, $%
\Omega _{C1}$ and $\Omega _{C2}$ [see Fig. \ref{fig1}(a)], may be neglected.
In this work, {we consider the density-distribution profile in Fig. \ref%
{fig1}(b) corresponding to $R(x)=\cos ^{2}x$,} i.e., $V(x)=V_{0}\cos ^{2}x$,
$C(x)=C_{0}\cos ^{2}x$, the notation being fixed by scaling the modulation
period to be $\pi $. Generic results are displayed below for amplitude $%
V_{0}=0.5$ of the nonlinearity modulation, while strength $C_{0}$ of the
linear coupling is varied. As for the sign of $C_{0}$, it may be fixed to be
positive, as $C_{0}<0$ can be transformed into $C_{0}>0$ by the change of $%
\phi \rightarrow -\phi $, while the sign of $\psi $ is not altered.

Stationary soliton solutions to Eqs. (\ref{eqs}) were found in a numerical
form by means of the imaginary-time propagation method \cite{Chiofalo}, and
their stability was subsequently tested by direct simulations of the
perturbed evolution in real time. The solitons are characterized by the
total power,
\begin{equation}
P\equiv P_{1}+P_{2}=\int_{-\infty }^{+\infty }\left[ |\psi (x)|^{2}+|\phi
(x)|^{2}\right] dx.  \label{power}
\end{equation}

\begin{figure}[tbp]
\centering%
\subfigure[] {\label{fig_4_a}
\includegraphics[scale=0.28]{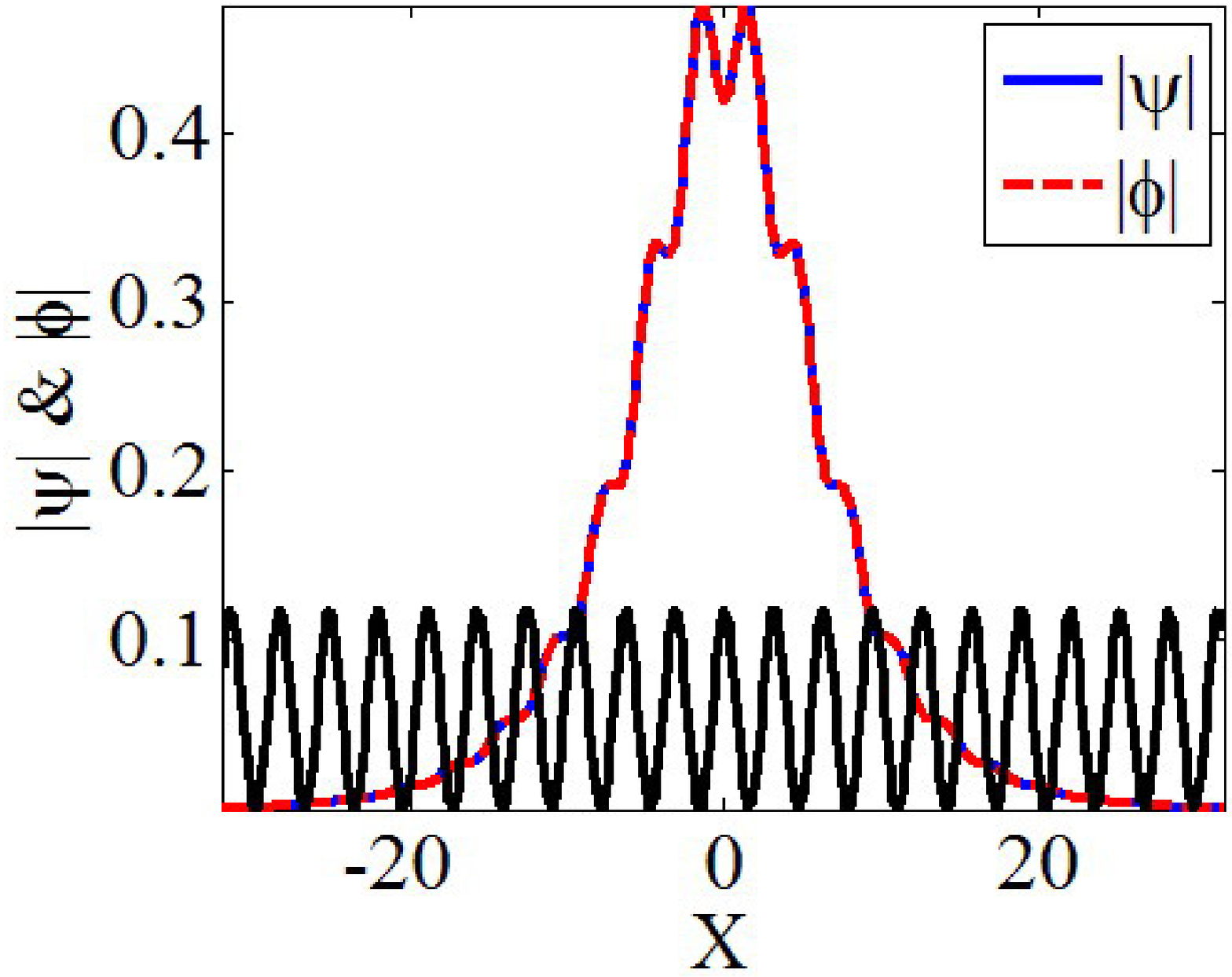}}%
\subfigure[] {\label{fig_3_a}
\includegraphics[scale=0.28]{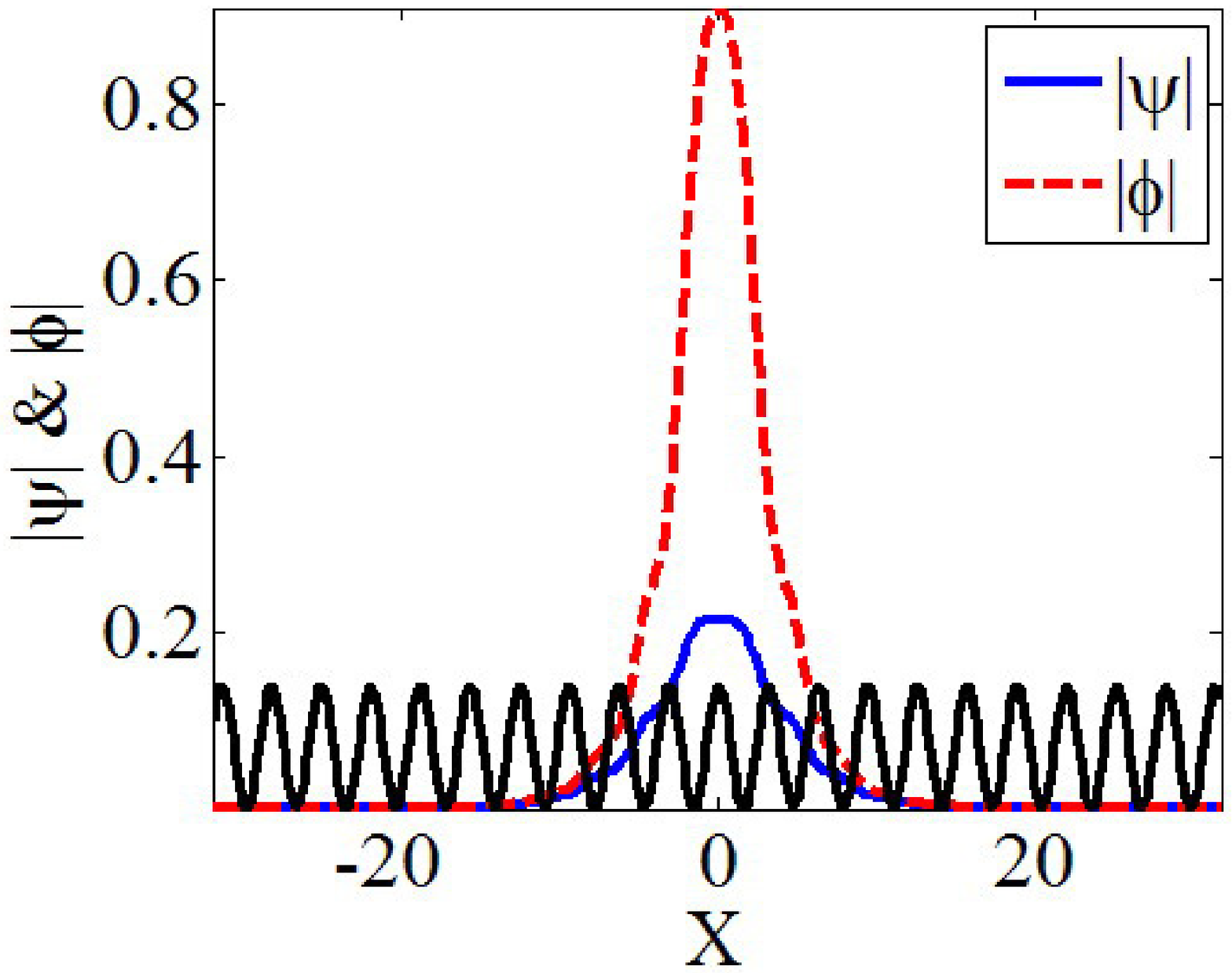}} %\subfigure[]{ \label{fig_3_b}
%\includegraphics[scale=0.28]{3b.eps}}
%\subfigure[]{ \label{fig_3_c}
%\includegraphics[scale=0.28]{3c.eps}}
\caption{(Color online) Stationary profiles of the two components of stable
symmetric (a) and asymmetric (b) solitons, both found for $P=4$, $C_{0}=0.078
$. These solitons belong to the bifurcation diagram displayed in Fig.
\protect\ref{fig_2}(a). In this figure and below,{\ the black solid lines
display the shape of the underlying modulation function, $R(x)$ [see Fig.
\protect\ref{fig1}(b)].}}
\label{fig_3}
\end{figure}

\begin{figure}[tbp]
\centering%
\subfigure[] {\label{fig_6_a}
\includegraphics[scale=0.28]{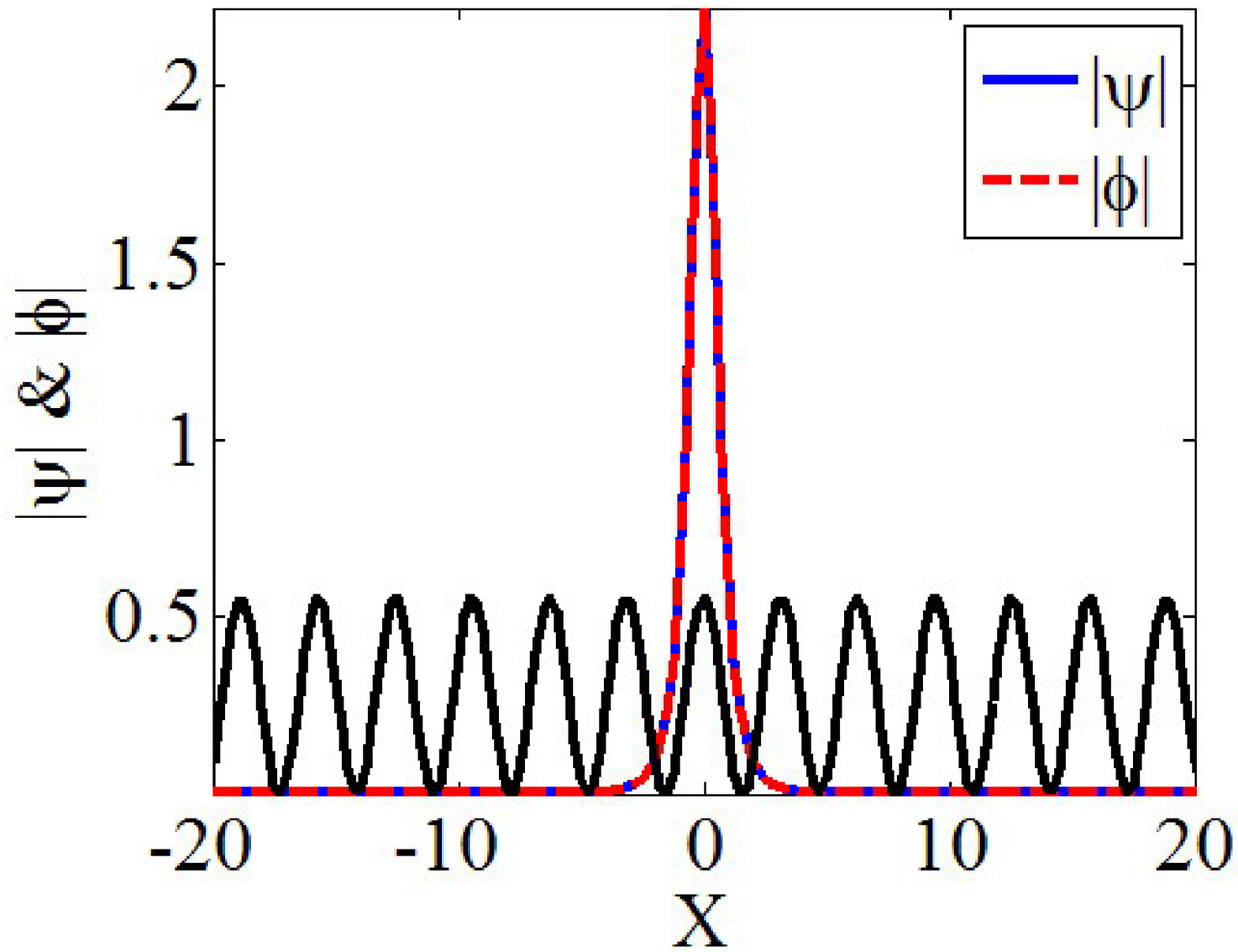}}%
\subfigure[] {\label{fig_5_a}
\includegraphics[scale=0.28]{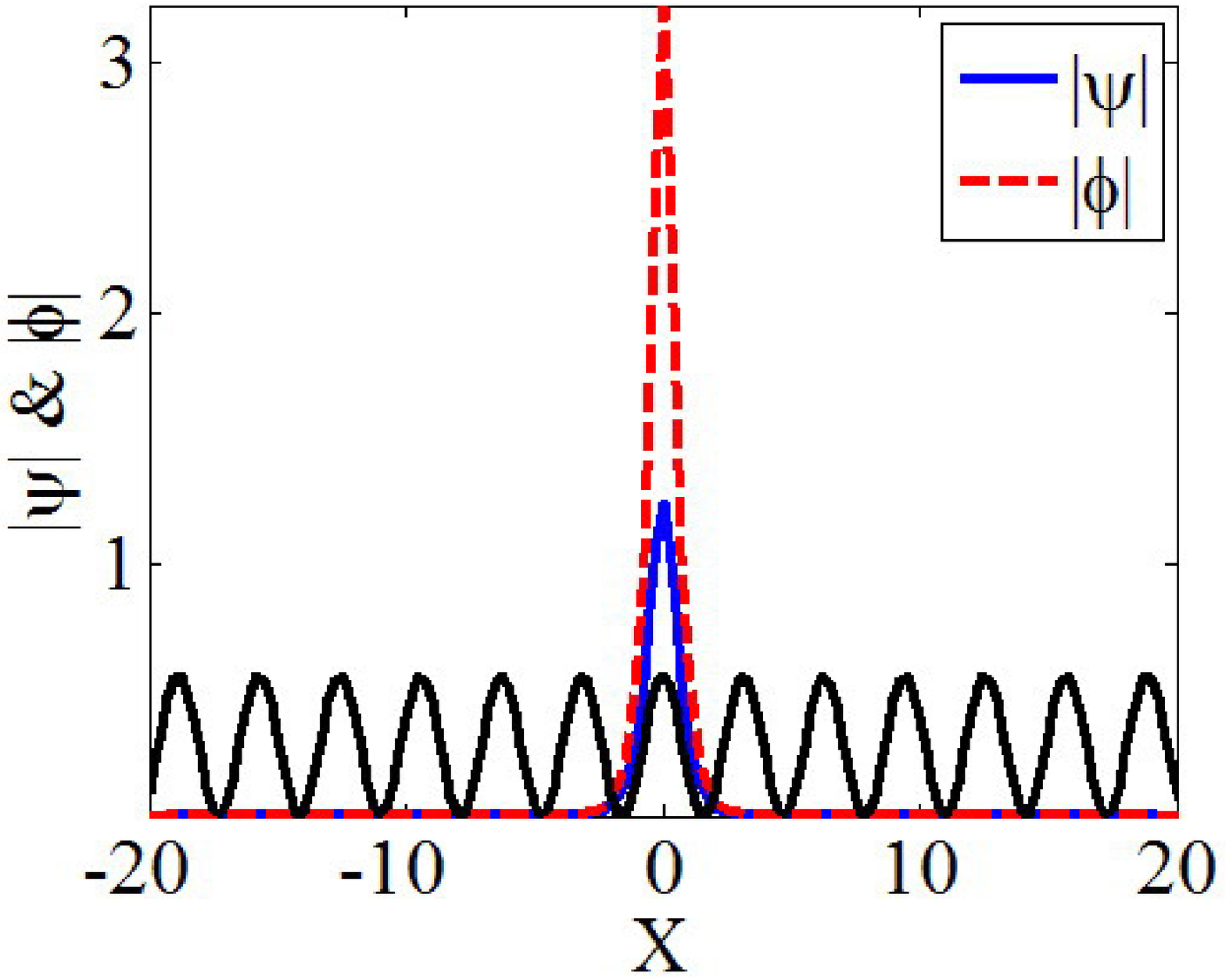}} %\subfigure[]{ \label{fig_5_b}
%\includegraphics[scale=0.28]{5b.eps}}
%\subfigure[]{ \label{fig_5_c}
%\includegraphics[scale=0.28]{5c.eps}}
\caption{(Color online) Examples of stable symmetric (a) and asymmetric (b)
solitons found for $P=10$ and $C_{0}=2$ (a) or $C=1.5$ (b). These solitons
belong to the bifurcation diagram displayed in Fig. \protect\ref{fig_2}(d).}
\label{fig_4}
\end{figure}

As expected, both symmetric and asymmetric soliton modes, in terms of the
coupled components, were found, see Figs. \ref{fig_3} and \ref{fig_4}. These
examples display stable solitons, which are broad in comparison with the
underlying modulation pattern at smaller values of $P$ [Fig. \ref{fig_3}],
and narrower modes, with the width comparable to the modulation period, at
larger $P$ [Fig. \ref{fig_4}].

%\begin{figure}[tbp]
%\centering
%\subfigure[]{ \label{fig_4_b}
%\includegraphics[scale=0.28]{4b.eps}}
%\subfigure[]{ \label{fig_4_c}
%\includegraphics[scale=0.28]{4c.eps}}
%\caption{(Color online) (a) $P=4$, $C_{0}=0.078$, the soliton solution are
%both located at the branches of Fig. (\protect\ref{fig_2_a}). }
%\label{fig_4}
%\end{figure}

%\begin{figure}[tbp]
%\centering
%\subfigure[]{ \label{fig_6_b}
%\includegraphics[scale=0.28]{6b.eps}}
%\subfigure[]{ \label{fig_6_c}
%\includegraphics[scale=0.28]{6c.eps}}
%\caption{(Color online) $P=10$, $C_{0}=1.5$, (a) the solution which is
%located at the center line of Fig. (\protect\ref{fig_2_d}). stable. }
%\label{fig_6}
%\end{figure}

\begin{figure}[tbp]
\centering
\subfigure[]{ \label{fig_2_a}
\includegraphics[scale=0.25]{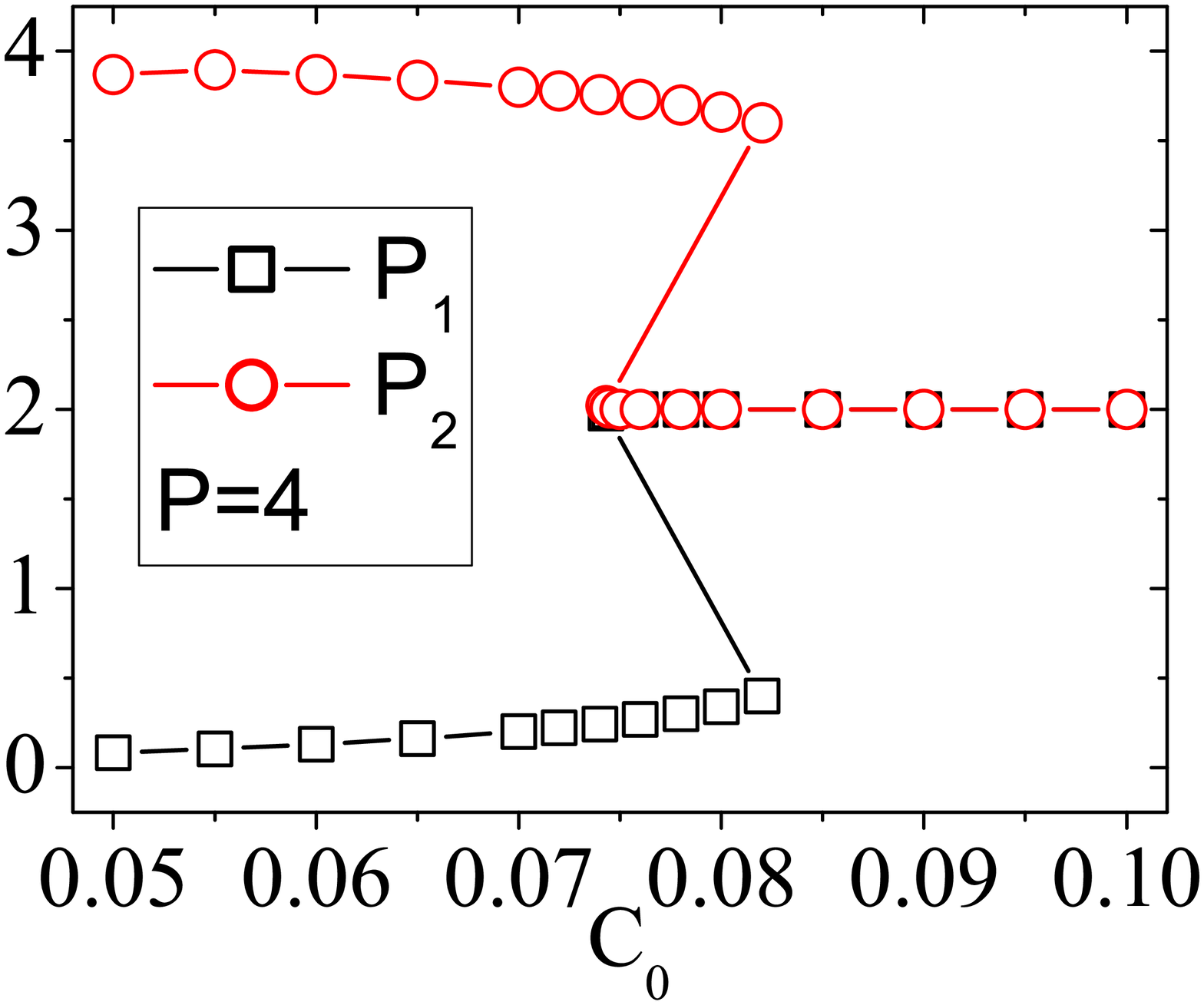}}
\subfigure[]{ \label{fig_2_b}
\includegraphics[scale=0.25]{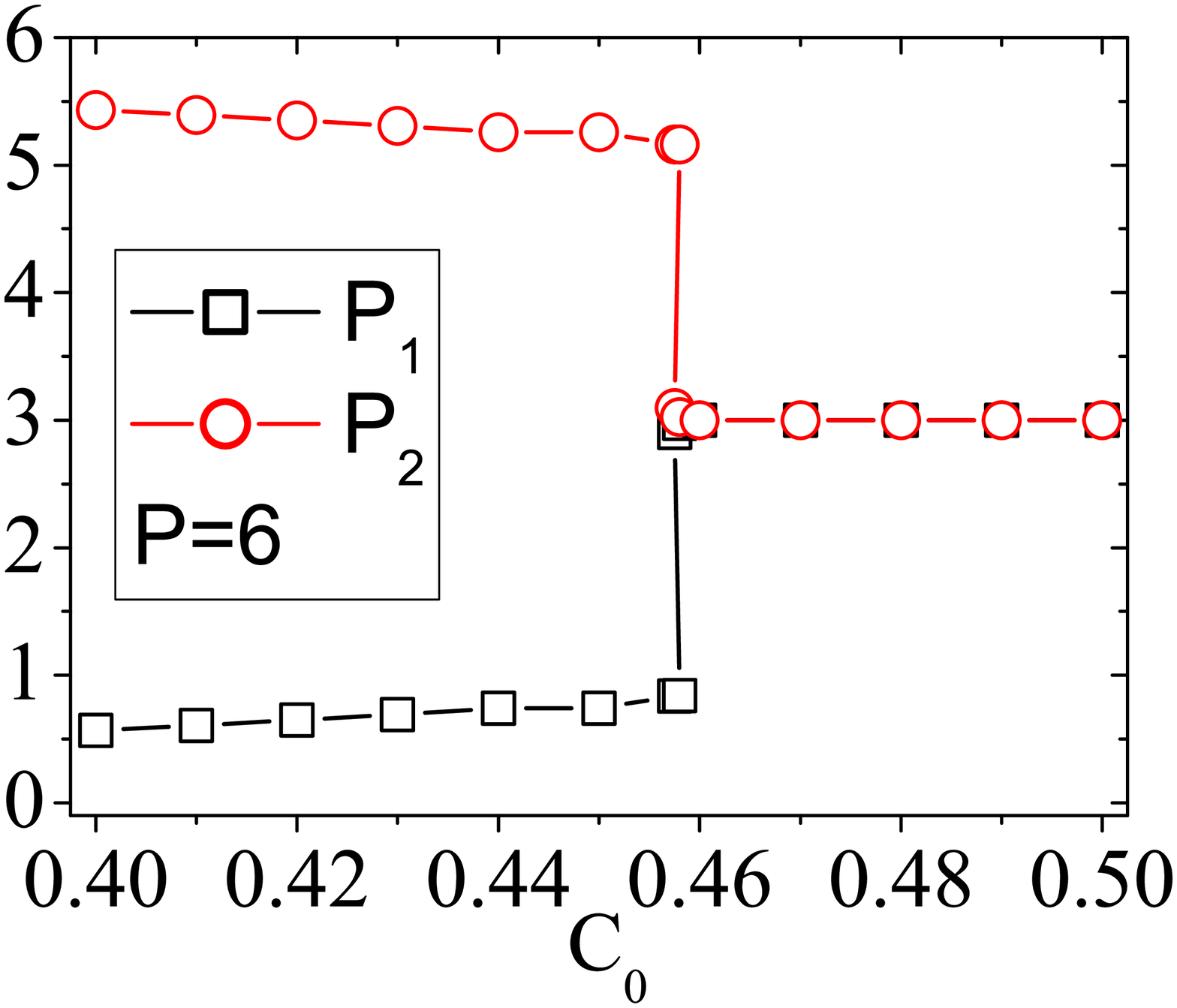}}
\subfigure[]{ \label{fig_2_c}
\includegraphics[scale=0.25]{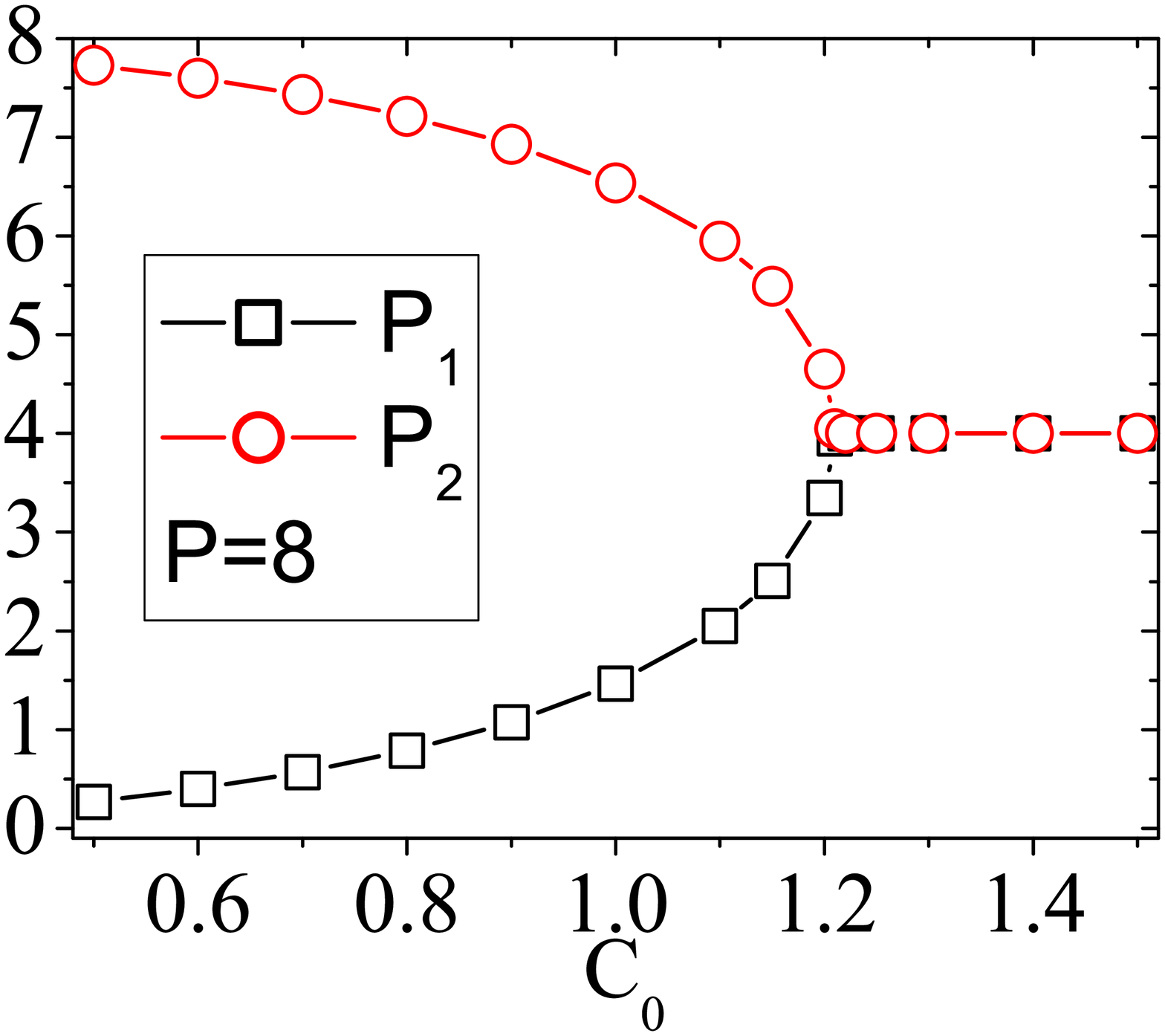}}
\subfigure[]{ \label{fig_2_d}
\includegraphics[scale=0.25]{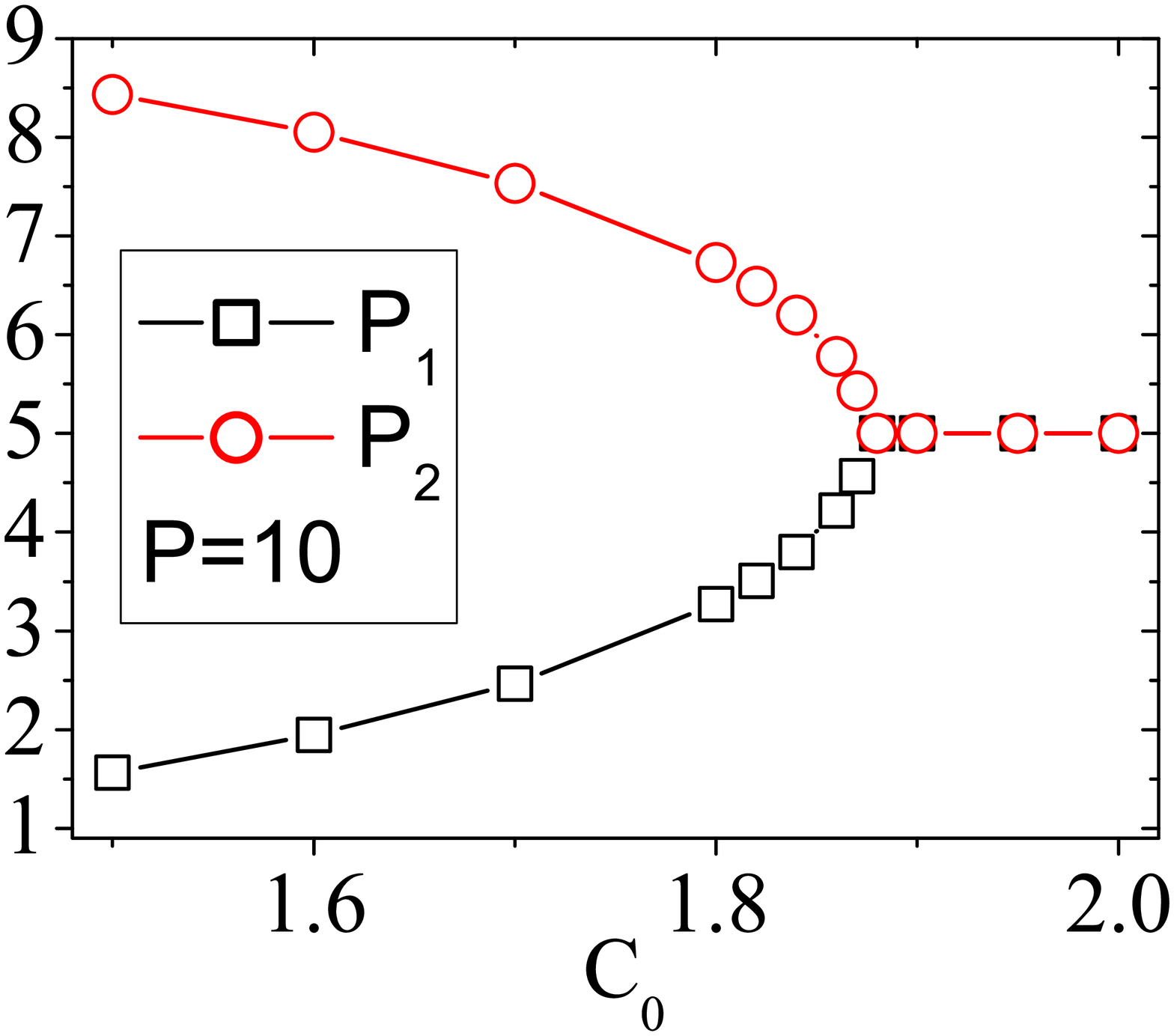}}
\caption{(Color online) The bifurcation diagrams for the two-component
solitons with different fixed values of the total power, corresponding to
the variation of strength $C_{0}$ of the linear coupling. Panels (a,b) and
(c,d) correspond to the bifurcations of the sub- and super-critical types,
respectively.}
\label{fig_2}
\end{figure}

The numerical results are summarized in the form of bifurcation diagrams
displayed in Fig. \ref{fig_2} at different fixed values of the total power, $%
P$. The bifurcations are driven by the decrease of the coupling coupling, $%
C_{0}$. Unstable branches of symmetric solitons, which should continue the
stable ones in all the panels, and narrow intermediate branches of unstable
asymmetric solitons in panel \ref{fig_2}(a), are missing as the
imaginary-time integration method does not converge to unstable solutions.
The diagrams show the transition from the subcritical SBB to the
supercritical bifurcation with the increase of $P$.

For broad solitons one may replace $R(x)=\cos ^{2}x$ in Eqs. (\ref{eqs}) by
its mean value, $\left\langle \cos ^{2}x\right\rangle =1/2$, which reduces
the system to the one studied in earlier works \cite{Snyder}, where the SBB
is subcritical, in accordance with Fig. \ref{fig_2}(a). The new situation is
actually found here for narrow solitons, for which the SBB turns out to be
supercritical.

\begin{figure}[tbp]
\centering
\subfigure[]{ \label{fig_7_a}
\includegraphics[scale=0.25]{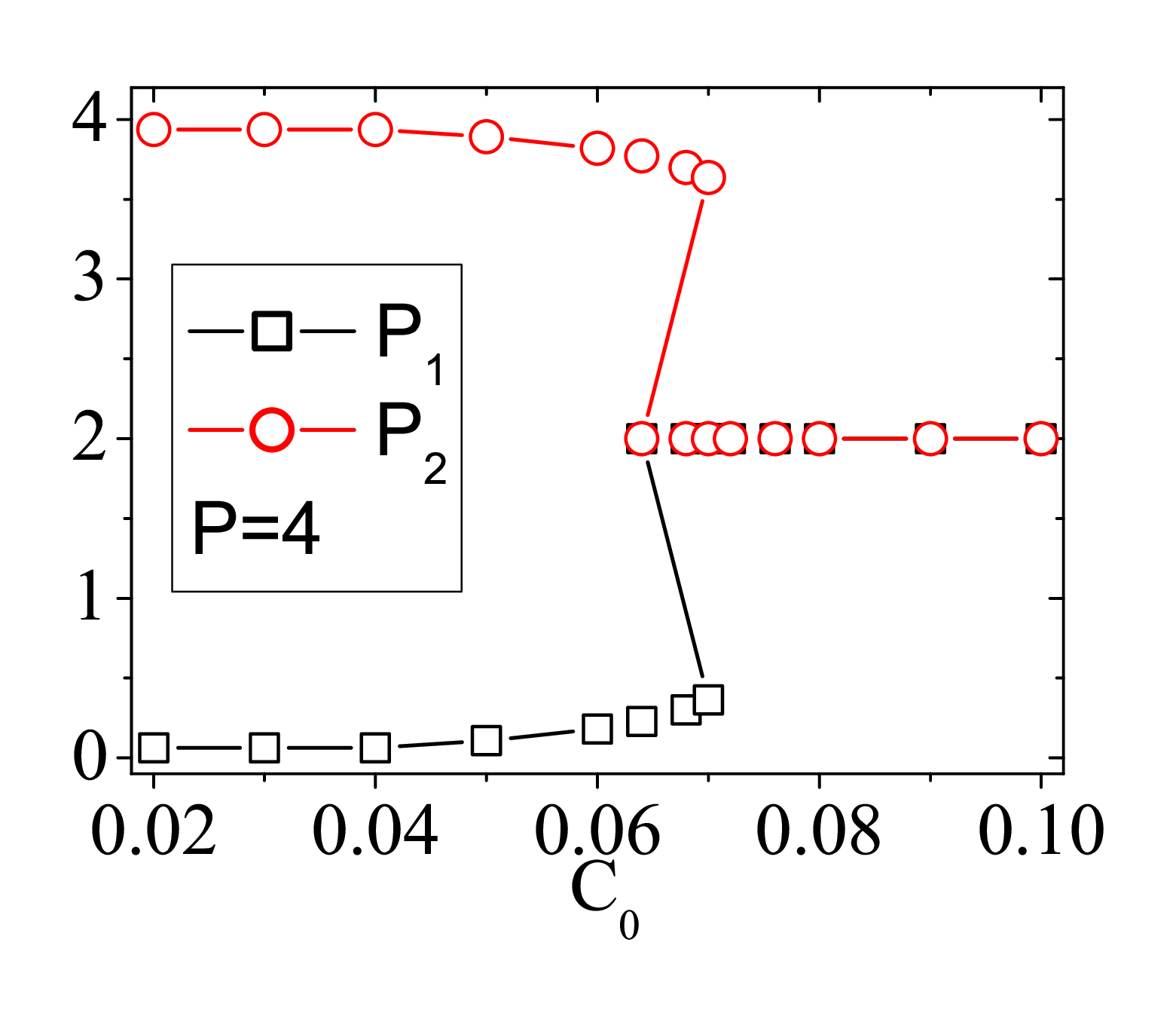}}
\subfigure[]{ \label{fig_7_b}
\includegraphics[scale=0.25]{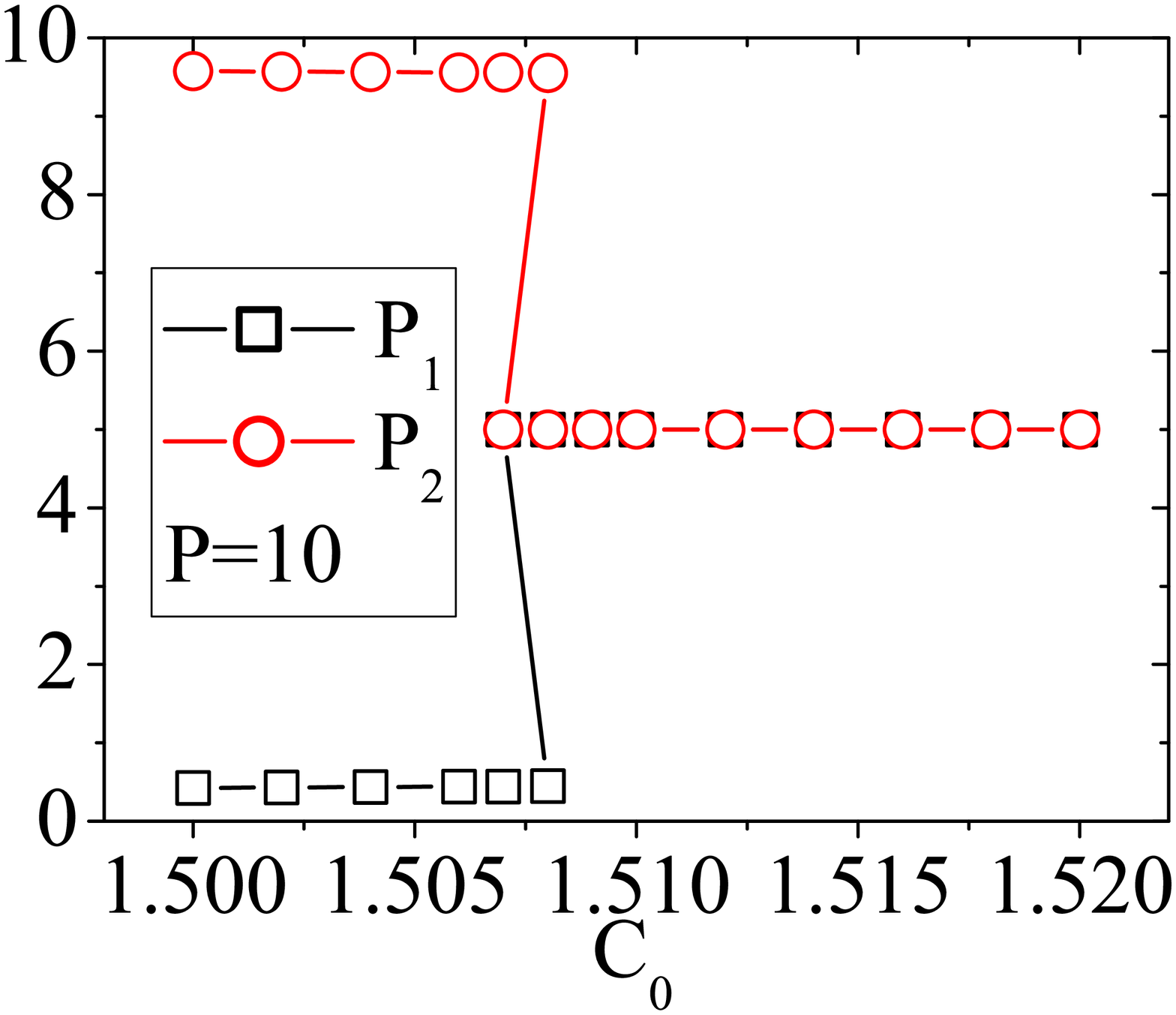}}
\caption{(Color online) The bifurcation diagrams for the two-component
solitons produced by modified system (\protect\ref{modified}).}
\label{fig_7}
\end{figure}

It is instructive to compare these results with those found in a system
where the linear and nonlinear potentials are the same as in Eqs. (\ref{eqs}%
), but the linear coupling is made constant, by replacing $C_{0}\cos
^{2}(x)\rightarrow C_{0}/2$:
\begin{eqnarray}
&&i\partial _{z}\psi =-{\frac{1}{2}}\partial _{xx}\psi +V(x)\left( 1-|\psi
|^{2}\right) \psi -\frac{1}{2}{C_{0}}\phi ,  \notag \\
&&i\partial _{z}\phi =-{\frac{1}{2}}\partial _{xx}\phi +V(x)\left( 1-|\phi
|^{2}\right) \phi -\frac{1}{2}{C_{0}}\psi .  \label{modified}
\end{eqnarray}%
Figure \ref{fig_7} demonstrates that the modified system does not give rise
to the transition of the subcritical SBB into the supercritical type, even
in the case when $P$ is large and the solitons are narrow. Thus, the
periodic modulation of the linear coupling is essential for this transition.

\section{Analytical results for the fused coupler}

Numerical finding presented in the previous section suggest that the spatial
modulation of the linear coupling is a crucial factor which determines the
kind of the phase transition (SBB) in the system, and the possible change of
the kind. In this section, we aim to illustrate the genericity of this rule
by means of exact analytical results, obtained in an allied system which
describes a \textit{fused} dual-core nonlinear planar waveguide \cite{fused}%
, with the linear coupling tightly concentrated at $x=0$:%
\begin{eqnarray}
i\partial _{z}\psi  &=&-\frac{1}{2}\partial _{xx}\psi -\left\vert \psi
\right\vert ^{2}\psi -2\delta (x)\phi ,  \notag \\
i\partial _{z}\phi  &=&-\frac{1}{2}\partial _{xx}\phi -\left\vert \phi
\right\vert ^{2}\phi -2\delta (x)\psi ,  \label{psiphi}
\end{eqnarray}%
where $\delta (x)$ is the delta-function. This system is a limit case of
Eqs. (\ref{eqs}), with $C(x)=2\delta (x)$ and $V(x)\equiv 1$ (in that case,
the constant linear potential is trivial in the present case and may be
dropped). Coefficients in the system can be fixed as in Eqs. (\ref{psiphi})
by means of an obvious rescaling. It is relevant to mention that symmetric
and asymmetric solitons in a discrete version of this system were recently
studied in Ref. \cite{Belgrade}, but the discrete system does not admit
exact solutions. In fact, the underlying system of Eqs. (\ref{eqs}) may also
be realized in terms of the dual-core spatial coupler, with periodically
modulated strength of the coupling, refractive index, and Kerr coefficient.

Stationary solutions to Eqs. (\ref{psiphi}) are sought for as $\left\{ \psi
,\phi \right\} =\exp \left( ikz\right) \left\{ u(x),v(x)\right\} ,$ with
real functions $u(x)$ and $v(x)$ satisfying equations%
\begin{eqnarray}
ku &=&\frac{1}{2}u^{\prime \prime }+u^{3}+2\delta (x)v,  \notag \\
kv &=&\frac{1}{2}v^{\prime \prime }+v^{3}+2\delta (x)u.  \label{uv}
\end{eqnarray}%
Obviously, solutions to Eqs. (\ref{uv}) are subject to the following
boundary conditions, produced by the integration in an infinitesimal
vicinity of $x=0$:%
\begin{eqnarray}
u^{\prime }\left( x=+0\right) -u^{\prime }\left( x=-0\right) &=&-4v\left(
x=0\right) ,~  \notag \\
v^{\prime }\left( x=+0\right) -v^{\prime }\left( x=-0\right) &=&-4u\left(
x=0\right) .  \label{bc}
\end{eqnarray}%
Exact soliton solutions to Eqs. (\ref{uv}) are looked for as%
\begin{equation}
\left\{ u(x),v(x)\right\} =\sqrt{2k}\mathrm{sech}\left( \sqrt{2k}\left(
|x|+\left\{ \xi ,\eta \right\} \right) \right)  \label{xieta}
\end{equation}%
with $\xi ,\eta >0$. Symmetric and asymmetric solutions correspond to $\xi
=\eta $ and $\xi \neq \eta $, respectively (the present model does not admit
antisymmetric solutions). Total power (\ref{power}) of solutions (\ref{xieta}%
) is%
\begin{equation}
P=2\sqrt{2k}\left[ 2-\tanh \left( \sqrt{2k}\xi \right) -\tanh \left( \sqrt{2k%
}\eta \right) \right] .  \label{P}
\end{equation}

The substitution of expressions (\ref{xieta}) into Eqs. (\ref{bc}) yields
the following equations which determine positive constants $\xi $ and $\eta $%
:%
\begin{equation}
\sqrt{\frac{k}{2}}\sinh \left( \sqrt{2k}\left\{ \xi ,\eta \right\} \right)
\mathrm{sech}^{2}\left( \sqrt{2k}\left\{ \xi ,\eta \right\} \right) =\mathrm{%
sech}\left( \sqrt{2k}\left\{ \eta ,\xi \right\} \right) .  \label{sinh}
\end{equation}%
Using notation $t_{1,2}\equiv \tanh \left( \sqrt{2k}\left\{ \xi ,\eta
\right\} \right) $, Eqs. (\ref{sinh}) can be transformed into the following
form:%
\begin{eqnarray}
\left( t_{1}-t_{2}\right) \left( t_{1}^{2}+t_{2}^{2}+t_{1}t_{2}-1\right)
&=&0,  \label{1} \\
t_{1}t_{2} &=&2/k,  \label{2}
\end{eqnarray}%
where physical solutions are constrained to $0<t_{1,2}<1$. Equation (\ref{1}%
) gives two solutions: $t_{1}=t_{2}$, which corresponds to symmetric
solitons, and the other solution, which corresponds to asymmetric ones:
\begin{equation}
t_{1}^{2}+t_{2}^{2}+t_{1}t_{2}=1.  \label{t12}
\end{equation}%
As it follows from Eq. (\ref{2}), the symmetric solutions have%
\begin{equation}
\tanh \left( \sqrt{2k}\left\{ \xi ,\eta \right\} \right) =\sqrt{2/k},
\label{sqrt}
\end{equation}%
hence they exist only for $k>2$, the total power (\ref{P}) of the symmetric
soliton being%
\begin{equation}
P_{\mathrm{symm}}=4\sqrt{2}\left( \sqrt{k}-\sqrt{2}\right) .  \label{Psymm}
\end{equation}

\begin{figure}[tbp]
\centering{\ \label{fig_8_a} \includegraphics[scale=0.4]{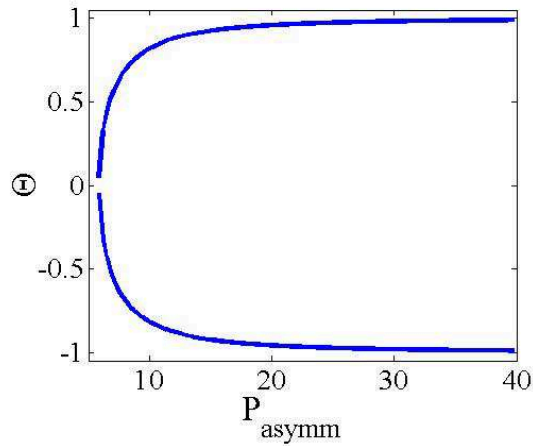}}
\caption{(Color online) Asymmetry $\Theta $ versus total power $P_{\text{%
asymm}}$, in the solvable model based on Eqs. (\protect\ref{psiphi}).}
\label{fig_8}
\end{figure}

The phase transition (SBB) occurs when the symmetric solution, $t_{1}=t_{2}=%
\sqrt{2/k}$, is simultaneously a solution to Eq. (\ref{t12}), which yields%
\begin{equation}
k=6,~\tanh \left( 2\sqrt{3}\left\{ \xi ,\eta \right\} \right) =\frac{1}{%
\sqrt{3}},~P_{\mathrm{symm}}=\frac{16}{\sqrt{3}}.  \label{bif}
\end{equation}%
The solution for the asymmetric solitons, following from Eqs. (\ref{t12})
and (\ref{2}), is%
\begin{equation}
\tanh ^{2}\left( \sqrt{2k}\left\{ \xi ,\eta \right\} \right) =\frac{k-2\pm
\sqrt{k^{2}-4k-12}}{2k},  \label{asymm}
\end{equation}%
which exists exactly at $k>6$, cf. Eq. (\ref{bif}). Further, the total power
(\ref{P}) of the asymmetric soliton is%
\begin{equation}
P_{\mathrm{asymm}}=2\left[ 2\sqrt{2k}-\left( \sqrt{k-2+\sqrt{k^{2}-4k-12}}+%
\sqrt{k-2-\sqrt{k^{2}-4k-12}}\right) \right] ,  \label{Pasymm}
\end{equation}%
and the relative asymmetry of the soliton is%
\begin{gather}
\Theta \equiv \frac{P_{1}-P_{2}}{P_{2}+P_{2}}=  \notag \\
\pm \frac{\sqrt{k-2+\sqrt{k^{2}-4k-12}}-\sqrt{k-2-\sqrt{k^{2}-4k-12}}}{2%
\sqrt{2k}-\left( \sqrt{k-2+\sqrt{k^{2}-4k-12}}+\sqrt{k-2-\sqrt{k^{2}-4k-12}}%
\right) }.  \label{Theta}
\end{gather}%
Figure \ref{fig_8} plots asymmetry $\Theta $ versus $P_{\text{asymm}}$, as
obtained from Eqs. (\ref{Theta}) and (\ref{Pasymm}). The figure confirms
that the SBB is indeed of the supercritical type when the linear coupling is
spatially localized.

The model based on Eqs. (\ref{psiphi}) is solvable too in the case of the
self-defocusing nonlinearity, corresponding to the opposite signs in front
of the cubic terms. However, in that case the model admits solely symmetric
solutions, in the form of $u(x)=v(x)=\sqrt{2k}/\mathrm{\sinh }\left( \sqrt{2k%
}\left( |x|+\xi \right) \right) $, with $\xi $ determined by equation $\tanh
\left( \sqrt{2k}\xi \right) =\sqrt{k/2}$, which has solutions for $k<2$, cf.
Eqs. (\ref{xieta}) and (\ref{sqrt}).

\section{Conclusions}

The objective of this work is to study the SBBs (symmetry-breaking
bifurcations) of solitons in two-component systems which, unlike previously
studied models, include the spatial modulation of the linear-coupling
strength. To this end, two photonic models were considered, namely, the
inverted virtual PhC (photonic crystal), and the fused dual-core spatial
coupler. The former system is built as the periodic distribution of the
density of dopant atoms, activated by the EIT, which induces the linear
mixing between the two probe fields. The periodic density modulation makes
all parameters of the medium periodic functions of the coordinate.
Disregarding the XPM terms, we have found that the type of the SBB changes
from sub- to supercritical with the increase of the total power of the probe
beams. In the model of the fused dual-core coupler, the solutions for the
two-component solitons were obtained in the exact form, the corresponding
SBB being supercritical.

The work can be naturally extended in other directions, including the
interplay with the spatial symmetry breaking, and the consideration of
higher-order solitons. A challenging possibility is to develop a
two-dimensional generalization of the system.

\begin{acknowledgments}
This work was supported by Chinese agencies NKBRSF (grant No. G2010CB923204)
and CNNSF(grant No. 11104083,10934011), by the German-Israel Foundation
through grant No. I-1024-2.7/2009, and by the Tel Aviv University in the
framework of the ``matching" scheme.
\end{acknowledgments}

%
%\bibitem{Kar_PO} Y. V. Kartashov, V. A. Vysloukh, and L. Torner, in Progress
%in Optics, edited by E. Wolf, Vol. 52 (North Holland, Amsterdam, 2009), p. 63.
%

%\newpage %Just because of unusual number of tables stacked at end
%

\bibliographystyle{plain}
\bibliography{apssamp}
% Produces the bibliography via BibTeX.

\end{document}